\documentclass[final]{raa}
\usepackage{graphicx,times}
\usepackage{natbib}
\usepackage{amssymb,amsmath}
\usepackage{textcomp}
\bibpunct{(}{)}{;}{a}{}{,}

\usepackage[a4paper=true,dvipdfm=true,pagebackref=true]{hyperref}
\hypersetup{pdftitle = The title of my PDF, pdfauthor = My name, pdfsubject= The subject, pdfkeywords = keyword1 keyword2 keyword3} 
\hypersetup{colorlinks = true, linkcolor = green, anchorcolor = red, citecolor = blue, filecolor = red, pagecolor = red, urlcolor = red}

\begin{document}

   \title{Discovery of Extended Structures around Two Evolved Planetary Nebulae M 2-55 and Abell 2}

   \volnopage{Vol.0 (2020) No.0, 000--000}      
   \setcounter{page}{1}          

   \author{Chih-Hao Hsia
      \inst{1}
   \and Yong Zhang
      \inst{2}
   \and Xi-Liang Zhang
      \inst{3}
   \and Tao Luo 
      \inst{1}
   }

   \institute{State Key Laboratory of Lunar and Planetary Sciences, Macau University of Science and Technology, Taipa, Macau, China; {\it chhsia@must.edu.mo} \\  
        \and
   School of Physics and Astronomy, Sun Yat-Sen University Zhuhai Campus, Tangjia, Zhuhai, 519082, China; {\it zhangyong5@mail.sysu.edu.cn} \\
        \and
             Yunnan Observatories, Chinese Academy of Sciences, Kunming 650216, China \\
   }

   \date{Received~~2019 xx xx; accepted~~2020~~xx xx}

\abstract{
We report a multi-wavelength study of two evolved planetary nebulae (PNs) M 2-55 and Abell 2. Deep optical narrow-band images ([\ion{O}{III}], 
H$\alpha$, and [\ion{N}{II}]) of M 2-55 reveal two pairs of bipolar lobes and a new faint arc-like structure. This arc-shaped filament 
around M 2-55 appears a well-defined boundary from southwest to southeast, strongly suggesting that this nebula is in interaction with its 
surrounding interstellar medium. From the imaging data of {\it Wide-field Infrared Survey Explorer} (WISE) all-sky survey, we discovered 
extensive mid-infrared halos around these PNs, which are approximately twice larger than their main nebulae seen in the visible. We also 
present a mid-resolution optical spectrum of M 2-55, which shows that it is a high-excitation evolved PN with a low electron density of 250 
cm$^{-3}$. Furthermore, we investigate the properties of these nebulae from their spectral energy distributions (SEDs) by means of archival 
data.
\keywords{infrared: ISM -- ISM: structure -- planetary nebulae: individual (M 2-55 and Abell 2) -- stars: AGB and post-AGB}
}

   \authorrunning{C. -H. Hsia et al.}            
   \titlerunning{Halos around M 2-55 and Abell 2}  

   \maketitle

\section{Introduction}           
\label{sect:introduction}

As descendants of asymptotic giant branch (AGB) stars, planetary nebulae (PNs) are important objects to understand galactic abundance 
distribution and the ending of stellar evolution of low- and intermediate-mass stars. When infrared (IR) technique was first developed, the 
PN NGC 7027 was found to have a strong excess in the IR, far above the continuum level expected from thermal free-free emission 
\citep{Gillett67}. This excess was soon identified as due to thermal dust emission. From the fact that PNs are descendants of AGB stars, 
\citet{Kwok82} predicted that the remaining circumstellar dust envelope from AGB stars should be commonly detectable in PNs. The prediction 
were confirmed by the {\it Infrared Astronomical Satellite} (IRAS) observations, where over 1000 PNs were detected \citep{Pottasch84}. The 
relative contributions from the stellar, nebular free-bound (f-b) and free-free (f-f) components, and dust continuum emissions in PNs have 
been analyzed by many researchers such as \citet{Zhang91} and \citet{Hsia14a}. 

The faint structures around PNs in the visible have been studied in several surveys (e.g. Stanghellini et al. 1993; Corradi et al. 2003; 
Cohen et al. 2011). Comparing to the observations at optical wavelengths, IR observations are less likely to be affected by the interstellar 
extinction. Thus it is highly desirable to observe faint structures around PNs in the IR. Moreover, unlike IR emissions, optical emissions 
are dominated by ionized gas. Therefore, a comparison study between IR and optical images allow us to better understand the processes of dust 
and gas components. Recent IR surveys have revealed new structures surrounding PNs \citep{Ramos09, Zhang12a, Zhang12b}, these results suggest 
that PNs may reveal different morphologies in the optical and IR. The {\it Wide-field Infrared Survey Explorer} (WISE) survey covers entire sky 
area and its sensitivity is 100 times higher than IRAS \citep{Wright10}. Thus, this survey provides a useful tool for us to resolve PNs with 
higher sensitivities \citep{Benjamin03, Churchwell09} and detect weak emissions from extended structures of known PNs. 

Evolved PNs represent the last stage of the dispersion of stellar material into the interstellar medium (ISM). \citet{Gurz69} first 
suggested that the interaction between PN and the ISM can decelerate the nebula. The observational evidence of PN-ISM interaction was later 
reported by \citet{Jacoby81}. The signs of PN-ISM interactions have been found to be common among evolved PNs \citep{Bork90}. To search and 
investigate the nebula/ISM boundary of evolved PNs can provide significant clue to study the mass-loss history of these objects and the matter 
enhancement in the galaxy.

The PNs M 2-55 (PN G116.2+08.5, IRAS 23296+7005) and Abell 2 (PN G122.1-04.9, IRAS 00426+5741) were first discovered and identified as evolved 
PNs by R. Minkowski in 1947 and \citet{Abell55} according to their low surface brightnesses and large angular diameters \citep{Abell66}. Based 
on early imaging studies, these objects have been classified as an irregular PN \citep{Felli79} for M 2-55 and an elliptical nebula 
\citep{Hua88} for Abell 2, respectively. Although their angular sizes ($\sim$60$\arcsec$) are relatively larger than most PNs ($\sim$
10$\arcsec$--30$\arcsec$), they are rarely paid attention in the past thirty years. After a keyword search in the astrophysics data system 
(ADS) for these objects, only four records are found in the literature. The distance determination of PNs has long been a difficult problem. 
Recently, trigonometric parallex data release (DR) of Gaia mission has induced an investigation of reliable distances for many PNs 
\citep{Bailer18}, allowing us to evaluate the distances of these two PNs with a higher accuracy.
 
In this paper, we present the results of a multi-wavelength investigation for PNs M 2-55 and Abell 2 based on optical and mid-infrared (MIR) 
images taken from {\it Lulin One-meter Telescope} and {\it WISE} all-sky survey, respectively, aiming to investigate the natures of the two 
PNs and their PN-ISM interactions. In $\S$ 2, we describe the observations and data reductions. The results of imaging and spectroscopic 
observations in the optical and MIR are presented in $\S$ 3. An analysis of the properties of the sources by analyzing their spectral energy 
distributions (SEDs) are presented in $\S$ 4. A discussion and conclusion are summarized in $\S$ 5 and $\S$ 6.

\section{Observations and data reduction}
\label{obs}

\begin{table*}
        \caption{Summary of Narrow-band Imaging Observations.}
        \label{tab1}
        \centering
        \begin{tabular}{ccccccc}
                \hline\hline\noalign{\smallskip}
                Object & Observation Date & Filter & $\lambda_{c}$ & $\triangle$$\lambda$ & Seeing & Exposures \\
                 &      &    &  (\AA~)  & (\AA~) & (arcsec) & (s) \\
                \hline\noalign{\smallskip}
 M 2-55 & 2018 Nov. 04 & H$\alpha$ & 6563 & 30 & 2.1 & 600$\times$20  \\
                & 2018 Nov. 05 & H$\alpha$ & 6563 & 30 & 1.7 & 1800$\times$11  \\
                & 2018 Nov. 07 & [\ion{O}{III}] & 5007 & 30 & 2.3 & 1800$\times$4  \\
            & 2018 Nov. 07 & [N II] & 6584 & 10 & 2.6 & 1800$\times$7  \\
 Abell 2 & 2018 Nov. 05 & H$\alpha$ & 6563 & 30 & 1.8 & 1800$\times$7  \\
                & 2018 Nov. 06 & H$\alpha$ & 6563 & 30 & 1.9 & 1800$\times$5  \\
                & 2018 Nov. 06 & [O III] & 5007 & 30 & 2.2 & 1800$\times$5  \\
                & 2019 Oct. 13 & [O III] & 5007 & 30 & 1.8 & 1800$\times$5  \\
                & 2019 Oct. 16 & [N II]& 6584 & 10 & 1.4 & 1800$\times$4  \\
                & 2019 Nov. 16 & [N II]& 6584 & 10 & 1.3 & 1800$\times$6  \\
\hline\hline\noalign{\smallskip}
        \end{tabular}
\end{table*}

\subsection{LOT Narrow-band Imaging}

Narrow-band Images of these nebulae were obtained from the {\it Lulin One-meter Telescope} (LOT) on the Lulin Observatory of National Central 
University (NCU) on the nights of 2018 November 4 -- 2019 November 16. The Lulin Compact Imager (LCI) Sophia 2048B camera with a 2048$\times$
2048 CCD was used. The CCD camera has a field of view (FOV) of 13.$\arcmin$2$\times$13.$\arcmin$2 with an angular resolution of 0.$\arcsec$39. 
These PNs were observed with three narrow-band filters: [\ion{O}{III}] ($\lambda_{c}$ = 5009 \AA, $\Delta\lambda$ = 30 \AA), H$\alpha$ 
($\lambda_{c}$ = 6563 \AA, $\Delta\lambda$ = 30 \AA), and [\ion{N}{II}] ($\lambda_{c}$ = 6584 \AA, $\Delta\lambda$ = 10 \AA). The total 
exposures were ranged from 7,200 to 31,800 s for each filter of the deep observations. We process all imaging data using the IRAF software 
package. Dark-current correction, bias subtraction, and flat-field calibration were performed. A summary of these observations is given in 
Table \ref{tab1}. The reduced [OIII], H$\alpha$, and [NII] images of these nebulae are shown in Figures~\ref{fig1} and ~\ref{fig2}, 
respectively.

\subsection{YFOSC Optical Spectra}

The spectra of M 2-55 were obtained on the nights of December 02 -- 03, 2018, with {\it Lijiang 2.4 m} telescope of the Yunnan Astronomy 
Observatories of China. The {\it Yunnan Faint Object Spectrograph and Camera} (YFOSC) instrument and a E2V 2048$\times$4608 CCD were 
used. The spectral dispersion of the spectra is $\sim$ 1.7 \AA~pixel$^{-1}$. The seeing conditions varied from 1.$\arcsec$7 and 2.$\arcsec$3 
during the observing runs. The wavelength coverage of spectroscopic observations is between 3600 and 7700~\AA. The aperture size of a slit 
is 5$\arcmin\times$ 1.$\arcsec$8 and it was set through the central region of this nebula oriented toward the N-S direction.
The exposures were ranged from 1800 to 11700 s and then the signal-to-noise (S/N) ratios of $>$ 75 for main nebula were produced. 

The data were reduced by a standard procedure using the NOAO IRAF V2.16 software package, including background subtraction, flat-fielding, 
and debiasing. For flux calibration, three KPNO standard stars each night were observed. To improve the S/N ratio, the final spectrum of this 
nebula was produced using the spectra with individual exposures. The journal of spectroscopic observations is given in Table \ref{tab2}.

Gaussian line profile fitting is used for measuring the fluxes of line emissions of this object. The uncertainties of line fluxes are 
derived from the continuum noise level. If we consider the flux errors of the measurements, the characteristic uncertainties of line 
emissions are about $\sim$5$\%$ - 37$\%$.

\begin{table*}
\caption{Summary of {\it YFOSC} Spectroscopic Observations.}
\label{tab2}
\centering
\begin{tabular}{cccccc}
\hline\hline\noalign{\smallskip}
Object & Observation Date & Wavelength & Dispersion & Width of Slit & Integration Time \\
       &                  &   (~\AA)    & (~\AA~pix$^{-1}$) & (arcsec) & (s) \\
\hline
M 2-55 & 2018 Dec 02 & 3600 - 7400 & 1.7 & 1.8 & 1800$\times$2  \\
       & 2018 Dec 03 & 3600 - 7400 & 1.7 & 1.8 & 2700$\times$3  \\
\hline\hline\noalign{\smallskip}
\end{tabular}
\end{table*}

\subsection{WISE Infrared Observations}

MIR images of these PNs were taken from the {\it WISE} all-sky survey. The WISE mission has imaged all sky with four bands at 3.4, 4.6, 12, 
and 22 $\mu$m (W1--W4). The angular resolutions of the images for these four bands are 6.$\arcsec$1, 6.$\arcsec$4, 6.$\arcsec$5, and 
12$\arcsec$, respectively. The data used in this paper were retrieved from the NASA/IPAC Infrared Science Archive (IRSA)
\footnote{http://irsa.ipac.caltech.edu/frontpage/.}.  

We performed the aperture photometric measurements for these nebulae using the method described in \citet{Hsia14b}. The fluxes measured 
from the WISE images of these PNs have calibrated using the color correction presented in \citet{Wright10}. To estimate the flux 
uncertainties, the standard deviations of systematic errors and all flux measurements are adopted. The estimated flux errors of the 
measurements in four bands for all objects are about 8$\%$ for W1, 9$\%$ for W2, 7$\%$ for W3, and 8$\%$ for W4 channels, respectively. 

\begin{figure}
\begin{center}
\includegraphics[width=0.8\textwidth]{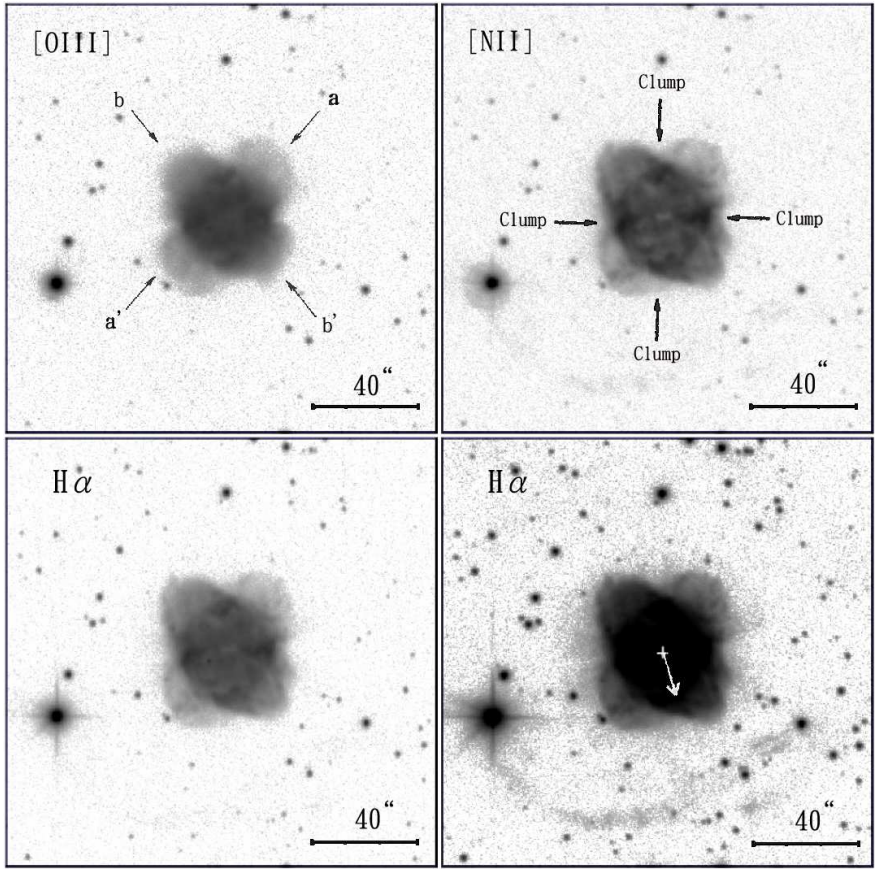}
\end{center}
\caption{Narrow-band images of M 2-55 in [\ion{O}{III}] (upper left), [\ion{N}{II}] (upper right), and H$\alpha$ (lower left and lower right) 
displayed with a logarithmic intensity scale. East is left and north is to the up. In the main nebula, two bipolar lobes (labeled as 
$a-a^\prime$ and $b-b^\prime$) and several clumps can be seen. Lower right panel: Same as on the lower-left panel but shown at different 
intensity levels. Outer extended filaments around this PN can be detected in the deep H$\alpha$ image. The position of central star is 
marked with cross. The white line denotes the direction of proper motion of this nebula.}
\label{fig1}
\end{figure}

\begin{figure}
\begin{center}
\includegraphics[width=0.8\textwidth]{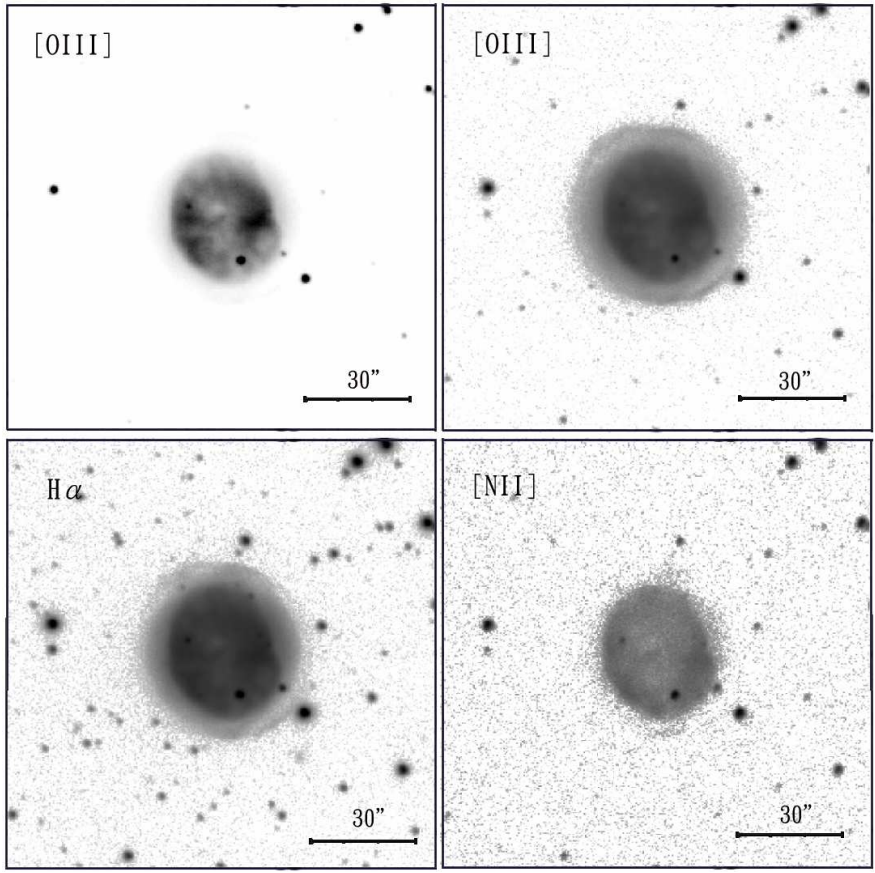}
\end{center}
\caption{(Upper left). Image of Abell 2 in [\ion{O}{III}] displayed with a linear scale, where few clumpy structures can be seen in the inner 
shell. The [\ion{O}{III}] (upper right), H$\alpha$ (lower left), and [\ion{N}{II}] (lower right) images of this object are displayed on a 
logarithmic intensity scale to show outer shell of the nebula. North is up and east is to the left.}
\label{fig2}
\end{figure}

\section{Results}
\label{result}

\subsection{Optical Morphology}


\subsubsection{Multipolar Planetary Nebula -- M 2-55}\label{M2-55}

The PN of M 2-55 is originally classified as a B-class nebula with symmetrical structures \citep{Sabb81}. Our {\it LOT} narrow-band images 
(H$\alpha$, [\ion{O}{III}], and [\ion{N}{II}]) of the nebula (Figure~\ref{fig1}) clearly show that this object is a multipolar PN with a size 
of $\sim$ 67.$\arcsec$6, which is larger than the size of 39$\arcsec$ measured from the optical image reported by \citet{Acker92}. The 
appearance of M 2-55 is similar to that of the young PN M 1-30 (see Figure~\ref{fig1} of Hsia et al. 2014) and can be related to the 
quadrupolar PN \citep{Manchado96}. The main structure of this PN consists two bipolar lobes (labeled as $a-a^\prime$ and $b-b^\prime$) and 
they are intersecting approximately at the central region of this object as shown in the H$\alpha$ and [\ion{N}{II}] images (Figure~\ref{fig1}). 

From Figure~\ref{fig1}, we can see that the edges of two bipolar lobes are sharp and they are prominent in the H$\alpha$ and [\ion{N}{II}] 
images. The central region of this nebula is mainly dominated by [\ion{O}{III}] emission, while the lobes are more obscure in the 
[\ion{O}{III}] image compared to those seen in the H$\alpha$ and [\ion{N}{II}] images. Several clumpy regions can be seen within the main 
nebula. These regions may be due to the projection of two interlaced lobes. We note that some ray-shaped structures and radial filaments 
pointing toward outer regions can be seen in the deep H$\alpha$ image. They may be the result of shadows of the clumpy region in the main 
nebula. These features can be also seen in other nebulae such as NGC 40, NGC 3242, NGC 3918, NGC 7009, NGC 7026, and NGC 7662 \citep{Corradi04}. 
The position angles (PAs) of two pairs of lobes ($a-a^\prime$ and $b-b^\prime$) are measured to be PA = 146$^\circ\pm$3$^\circ$ and 
42$^\circ\pm$3$^\circ$. We measured the sizes of two lobes by fitting the shapes of these features to deep H$\alpha$ image. Two pairs of 
bipolar lobes have approximately the same extent. The angular sizes of two bipolar lobes are measured to be 67.$\arcsec$2$\times$36.$\arcsec$6 
and 65.$\arcsec$5$\times$34.$\arcsec$3 for lobes $a-a^\prime$ and $b-b^\prime$, respectively.

Adopting a distance of 691 pc for M 2-55 \citep{Bailer18}, the extent size is 0.23 $\sec$$\theta$ pc for lobe $a-a^\prime$, where $\theta$ is 
the inclination angle. The expansion velocity is assumed to be 20.3 km s$^{-1}$ with [\ion{O}{III}] emission \citep{Wein89}, the derived 
kinematic age of this nebula is about 5500 $\sec$$\theta$ yr. This estimation is in good agreement with earlier suggestion that M 2-55 is an 
evolved PN \citep{Sati99}.    

\subsubsection{Arc-like Structure Discovered around M 2-55}\label{arc}

The filamentary appearance of the halo as an observational evidence can provide that the nebula interacts with high-density ISM \citep{Dgani98, 
Ali12}. As a PN with low density moving through its surrounding ISM, the leading region of the halo is compressed and then the 
arc-like/bow-shock structure forms \citep{Ali12}.         

\begin{figure}
\begin{center}
\includegraphics[width=0.8\textwidth]{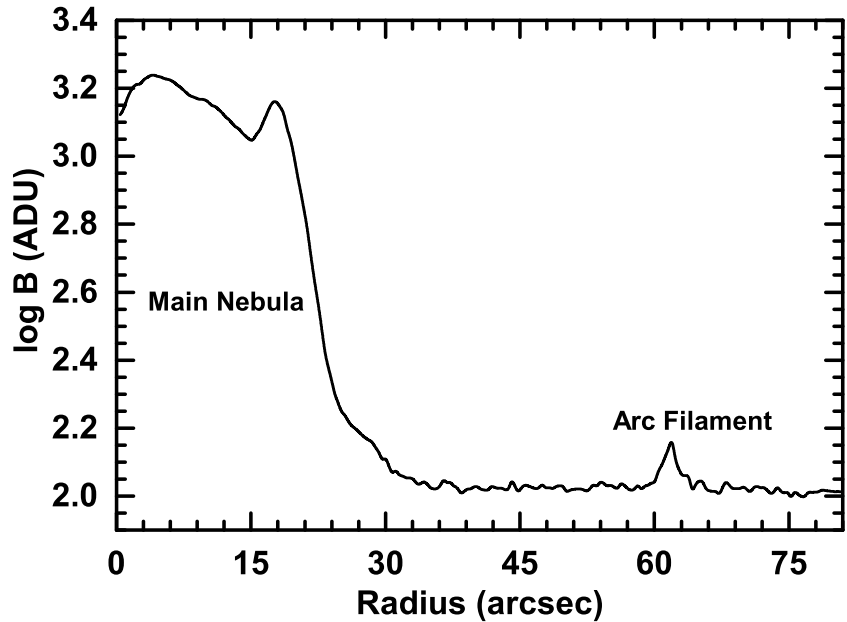}
\end{center}
\caption{H$\alpha$ surface brightness profile in M 2-55 averaged over the angles between PA = 155$^\circ$ and PA = 185$^\circ$. Horizontal
axis is the distance from the central star in unit of arcsec. The position of arc-shaped structure is marked.}
\label{fig3}
\end{figure}

A new arc-like filament extending to $\sim$61$\arcsec$ from the central star of this nebula can be seen in our deep H$\alpha$ image (see 
Figure~\ref{fig1}). The filamentary feature has not been detected in previous optical images. This arc shows a well-defined boundary from 
SW to SE, indicating that it may be a bow-shocked structure. The existence of this arc-shaped structure may suggest the PN-ISM interaction 
\citep{Ramos18, Wareing07}, which is supported by the proper motion direction of M 2-55 with PA = 192$^\circ$ \citep{Roeser10}. The surface 
brightness (SB) profile of this filamentary feature with H$\alpha$ emission is shown in Figure~\ref{fig3}. We made this profile from the 
average of the intervals between PA = 155$^\circ$ and PA = 185$^\circ$, after removing all field stars. From Figure~\ref{fig3}, the 
peak surface brightness of arc-shaped structure (at the distance of 61$\arcsec$ from the central star) is $\sim$ 6$\times$10$^{-2}$ times 
fainter than that of main nebula, and the feature is significantly brighter than normal AGB halo (with a SB 10$^{-3}$ times fainter than main 
shell; Corradi et al. 2003). Adopting a distance of 691 pc \citep{Bailer18} and assuming a typical expansion velocity of 13.7 km s$^{-1}$ for 
this faint structure \citep{Guss94}, the observed size of this extended structure leads to a dynamical age of $\sim$ 15,000 yr. This suggests 
that the arc-like feature can be an important event of its AGB mass-loss history.

\subsubsection{Double-shell Planetary Nebula -- Abell 2}

As can be seen in our narrow-band imaging, some structural features of PN Abell 2 show slightly different appearances depends on observed bands 
(Figure~\ref{fig2}). A close inspection of Figure~\ref{fig2} shows that two concentric elliptical shells (a brighter inner shell and a diffuse 
outer shell) are located in the main nebula. The limb-brightening inner shell has a size of $\sim$37.$\arcsec$4$\times$31.$\arcsec$2 and it 
is surrounded by a diffuse outer shell with a diameter of $\sim$52.$\arcsec$3$\times$47.$\arcsec$6, which is larger than previously reported 
size of 35$\arcsec\times$35$\arcsec$ \citep{Hua88}. The major axes of these shells are oriented along PA = 33$^\circ$. From Figure~\ref{fig2}, 
we can see that the inner elliptical shell with a clear boundary is prominent in H$\alpha$, [\ion{O}{III}], and [\ion{N}{II}] emissions, 
whereas the outer shell is undetectable in [\ion{N}{II}]. These results suggest that the inner shell of this nebula is relatively lower 
excitation and outer diffuse shell farther away is relatively higher excitation, which is unusual among common PNs \citep{Gue18}. 
Inside the inner shell of this nebula, a few small 
structural components can be seen in the [\ion{O}{III}] and H$\alpha$ images. The origin and formation of these structures are still unclear. 
It is possible that these features are produced from ionized non-uniform dense clumps. For this PN, the averaged value of nebular minor and 
major axial lengths as its size was adopted. At a distance of 2.82 kpc \citep{Bailer18}, the average nebular size of this source is 0.68 pc, 
we obtained the dynamical age of $\sim$ 9,800 yr. Assuming an expansion velocity of the nebula is 34 km s$^{-1}$ \citep{Acker92}. This 
suggests that this nebula is indeed an old PN \citep{Abell66, Hua88}.  
 
\subsection{Extended Dust Halos around M 2-55 and Abell 2}\label{halo}

\begin{figure}
\begin{center}
\includegraphics[width=0.9\textwidth]{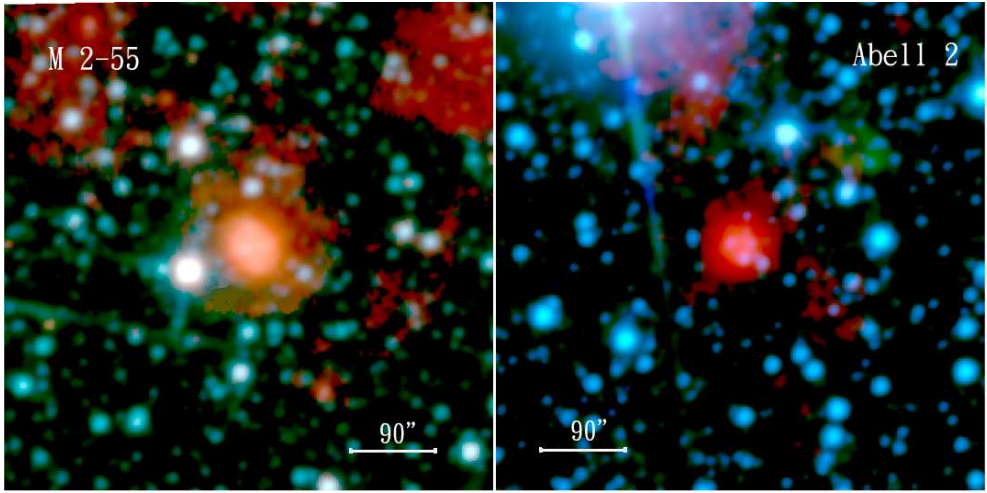}
\end{center}
\caption{Color composite {\it WISE} images of M 2-55 and Abell 2. The infrared images are shown on a logarithmic scale. These PNs are made
from three bands: 3.4 $\mu$m  (shown as blue), 4.6 $\mu$m (green), and 22 $\mu$m (red). North is up and east is to left. The extended
halos around these PNs can be seen in the images.}
\label{fig4}
\end{figure}

To study the dust distributions and their properties of PNs M 2-55 and Abell 2, we have analyzed the MIR images of these nebulae retrieved from 
the {\it WISE} all-sky survey archive. The extended halos with IR emission surrounding the objects can clearly be seen in Figure~\ref{fig4}. 
The MIR images show that the colors of the field stars are bluer than those of the nebulae, which suggests that they are dusty. The halos are 
detectable at {\it WISE} 12 and 22 $\mu$m bands, and they are brighter in the 22 $\mu$m images compared to the 12 $\mu$m images for these 
nebulae. Assuming that their emission peaks at 22 $\mu$m, we infer that IR fluxes emitted from the halos of these PNs are mainly dominated 
by dust components with a temperature of $\sim$130 K. From Figure~\ref{fig4}, we also note that PN M 2-55 shows lower IR surface brightness 
compared to PN Abell 2. 
In the {\it WISE} 12 and 22 $\mu$m images, the central parts of these nebulae show prominent IR emissions, probably suggesting a large number 
of dust located in the central regions of these PNs although the emissions of [\ion{Ne}{V}] $\lambda$24.3 $\mu$m and [\ion{O}{IV}] $\lambda$
25.89 $\mu$m may partly contribute to the fluxes of 22 $\mu$m band emissions \citep{Chu09}. 
    
The {\it WISE} color composite image of M 2-55 (left panel of Figure~\ref{fig4}) clearly shows that this PN has a slightly elliptical central 
nebula and an extended outer halo. The morphology of this nebula in the IR is different from that observed in the {\it LOT} optical narrow-band 
images. In this image, the central nebula is the brightest feature of this object, whereas the outer halo is 2$\times$10$^{-2}$ fainter. The 
southern part of the halo exhibits a well-defined boundary (with a size of $\sim$166$\arcsec$), in contrast to a diffuse structure seen in 
the northern part of the halo. Such morphology gives an indication that this source is moving roughly toward the south direction and probably 
shows the PN-ISM interaction \citep{Ramos09, Zhang12b}. For Abell 2, the optical and MIR morphologies of this object are similar. Our MIR 
image (right panel of Figure~\ref{fig4}) shows an almost round halo with an angular diameter of $\sim$92$\arcsec$. The central 
elliptical-shaped nebula is prominent in all {\it WISE} IR images. No evidence of related extended structure is found outside the halo of 
this object.  

\subsection{Spectral Properties of M 2-55}

In order to understand the spatial distributions of extended arc-shaped structure with H$\alpha$ emission around PN M 2-55 (see 
Figure~\ref{fig1}), we have carried out spectroscopic observations. However, the faint filamentary feature is undetectable in our 
spectroscopic observations due to its low surface brightness compared to main shells of this PN (see Section~\ref{arc}). 
Nevertheless, the properties of the main nebula can still be studied from our spectra. 

\begin{figure}
\begin{center}
\includegraphics[width=0.9\textwidth]{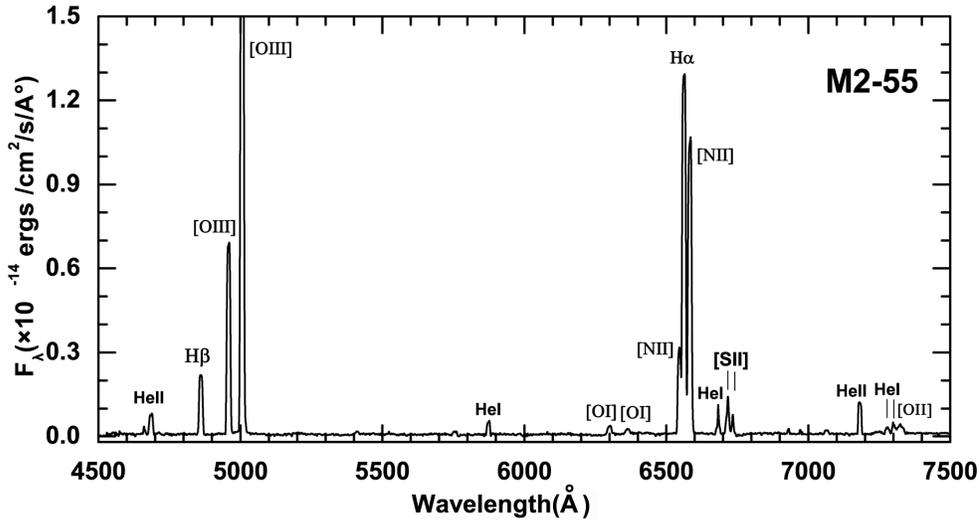}
\end{center}
\caption{Optical spectrum of M 2-55 in the wavelength range of 4500~\AA~to 7500~\AA. The prominent emission features are marked.}
\label{fig5}
\end{figure}

The YFOSC spectrum of PN M 2-55 is shown in Figure~\ref{fig5}. From Figure~\ref{fig5}, we can see that a number of emission lines are typical 
in PN spectra. Among the prominent emission lines detected are H$\beta$ at 4861~\AA, H$\alpha$ at 6563~\AA, [\ion{O}{III}] at 4959, 5007~\AA, 
[\ion{N}{II}] at 6548, 6584~\AA, and [\ion{S}{II}] at 6717, 6731~\AA. Some weak emission features such as \ion{He}{I} $\lambda\lambda$5876, 
6678, 7281, 7298, \ion{He}{II} $\lambda\lambda$4686, 7178, [\ion{O}{I}] $\lambda\lambda$6300, 6364, and [\ion{O}{II}] $\lambda\lambda$ 7320 
can also be seen. The N(He)/N(H) and log~N/O ratios \citep{Peimbert83} suggest that the nebula is a type I PN \citep{Peimbert83, Peimbert87}, 
which is believed to originate from massive progenitor star \citep{Calvet83, King94}. The H$\alpha$/H$\beta$ flux ratio measured from M 2-55 
is 5.95$\pm$0.48. Given the theoretical values at T$_{e}$ =10$^{4}$ K and n$_{e}$ = 10$^{2}$ cm$^{-3}$ \citep{Hummer87}, the extinction value 
of c = 1.13$\pm$0.13 is derived by comparing the observed H$\alpha$/H$\beta$ ratio and using the reddening law with a R$_{V}$ = 3.1 
\citep{How83}. This extinction is in good agreement with the earlier reported value of c = 1.24 \citep{Kaler90}.
      
\begin{table*}
	\caption{Characteristic Lines in M 2-55}
	\centering
	\begin{tabular}{lcccc}
		\hline\hline
		& \multicolumn{2}{c}{Identification} & \\
		\cline{2-3}
		$\lambda_{obs}$ & $\lambda_{lab}$ & Ion & Observed Flux$^a$ & Dereddened Flux$^a$ \\
		(\AA~)  &  (\AA~) &  &   &  \\
		\hline\noalign{\smallskip}
		4685.54 & 4685.68 & \ion{He}{II} & 4.41 (16.5) & 11.39 (16.5) \\ 
		4860.74 & 4861.33 & H$\beta$ & 16.81 (5.4) & 37.78 (5.4) \\ 
                4958.45 & 4958.91 & [\ion{O}{III}] & 44.45 (13.8) & 96.30 (13.8) \\
                5006.31 & 5006.84 & [\ion{O}{III}] & 137.91 (7.8) & 289.80 (7.9) \\
		5874.36 & 5875.66 & \ion{He}{I} & 2.35 (13.6) & 3.10 (13.7) \\
		6300.03 & 6300.34 & [\ion{O}{I}] & 2.20 (18.5) & 2.43 (18.5) \\
		6364.57 & 6363.78 & [\ion{O}{I}] & 1.11 (36.7) & 1.19 (36.7) \\
		6547.04 & 6548.10 & [\ion{N}{II}] & 17.24 (13.9) & 17.34 (13.9) \\
		6562.34 & 6562.77 & H$\alpha$ & 100.00 (3.2) & 100.00 (3.2) \\ 
		6582.81 & 6583.50 & [\ion{N}{II}] & 56.56 (4.7) &  56.12 (4.7) \\
		6679.13 & 6678.16 & \ion{He}{I} & 2.76 (5.4) & 2.64 (5.4) \\
		6716.07 & 6716.44 & [\ion{S}{II}] & 4.60 (3.7) & 4.35 (3.7) \\
		6730.39 & 6730.82 & [\ion{S}{II}] & 3.73 (7.6) & 3.51 (7.6) \\
		7178.15 & 7177.50 & \ion{He}{II} & 6.39 (5.0) & 5.17 (5.0) \\ 
		7280.68 & 7281.35 & \ion{He}{I} & 1.22 (11.1) & 0.96 (11.1) \\ 
		7299.17 & 7298.04 & \ion{He}{I} & 1.59 (8.9) & 1.24 (8.9) \\
		7321.46 & 7319.99 & [\ion{O}{II}] & 3.07 (7.1) & 2.38 (7.1) \\
		\hline\hline\noalign{\smallskip}
	\end{tabular}
	\tablecomments{0.86\textwidth}{[a] The normalized emission fluxes (H$\alpha$ = 100). The brackets represent the flux uncertain 
errors with percentage.}
	\label{tab3}
\end{table*} 

The measured emission fluxes are given in Table~\ref{tab3}. Columns 1 - 3 list the observed emission wavelengths and line identifications. 
The normalized measured and dereddened fluxes (H$\alpha$ = 100) of this object are listed in Columns 4 and 5. The observed integrated 
H$\beta$ flux measured from this PN is 3.73$\times$10$^{-14}$ erg cm$^{-2}$ s$^{-1}$. From Table~\ref{tab3}, the line ratios of [\ion{O}{III}] 
$\lambda$5007/$\lambda$4959 and [\ion{N}{II}] $\lambda$6584/$\lambda$6548 of this nebula are 3.0 and 3.2, respectively, which are consistent
with the theoretical values suggested by \citet{Storey00}. Adopting an electron temperature of $T_{e}$ = 10,200 K \citep{Peimbert87} for this 
nebula and using the [\ion{S}{II}] $\lambda$6731/$\lambda$6717 line ratio, the electron density $n_{e}$ = $250_{-190}^{+210}$\, cm$^{-3}$ was derived. This value is in good agreement with the earlier results of 460 cm$^{-3}$ \citep{Peimbert87} and 512 cm$^{-3}$ \citep{Phillips98}. 
The slight difference stems from different diaphragm sizes and slit positions.
The emission ratio of log([\ion{O}{III}] $\lambda$5007+$\lambda$4959/HeII $\lambda$4686) is an useful probe to determine the excitation class of a PN \citep{Gurz88}.
For this object, the value of log([\ion{O}{III}] $\lambda$5007+$\lambda$4959/HeII $\lambda$4686) ratio is about 1.53, which indicates that
this nebula is a high-excitation class PN. Moreover, the H$\alpha$/[\ion{O}{III}] $\lambda$5007 flux ratio of M 2-55 is 0.35, which is
smaller than that of normal young PNs ($>$1.5; Sahai \& Trauger 1998), suggesting that this object is evolved.

\section{Spectral Energy Distribution}\label{sed}

\begin{figure}
\begin{center}
\includegraphics[width=0.9\textwidth]{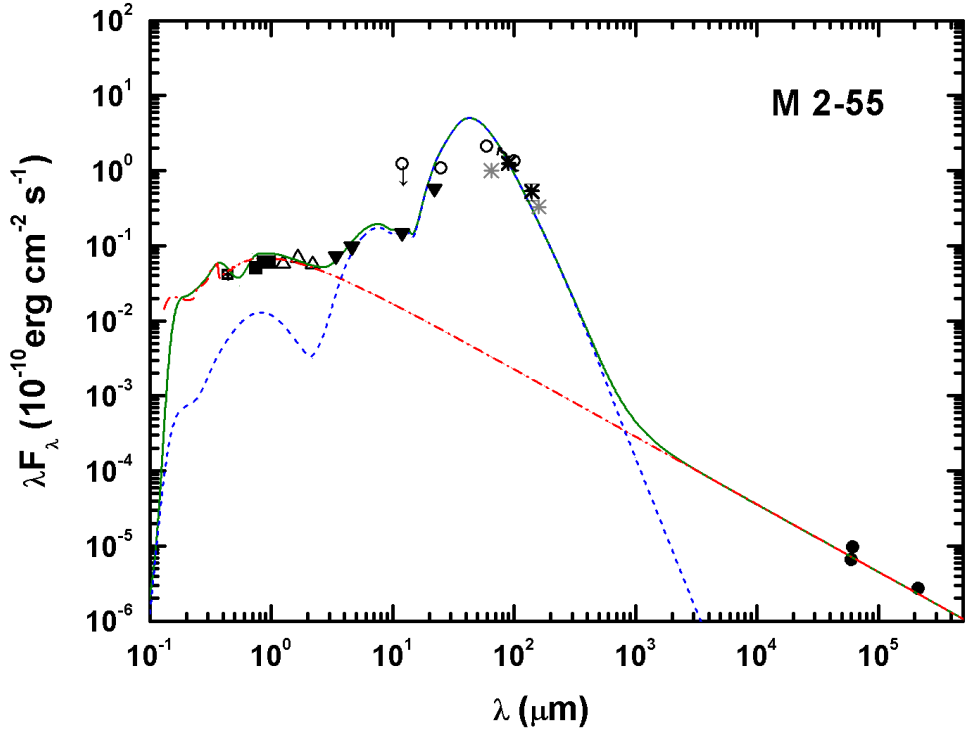}
\end{center}
\caption{The SED of M 2-55 with wavelength range from UV to radio. The Pan-Starrs measurements are shown as filled squares, the NUV, B,
and V measurements as open squares, the 2MASS results as open triangles, WISE measurements as filled triangles, IRAS as open circles, AKARI
photometry as asterisks, and radio measurements as filled circles. The uncertain AKARI detections are marked as the light asterisks. Note that
the measured IRAS 12 and 100 $\mu$m fluxes are upper limit detections. The ISO LWS spectrum for M 2-55 is also plotted. The nebular emissions 
are plotted as the red curve. The blue curve represents a dust continuum fitting by using the radiation transfer model. The total fluxes 
derived from all components are plotted as the green curve.}
\label{fig6}
\end{figure}

\begin{figure}
\begin{center}
\includegraphics[width=0.9\textwidth]{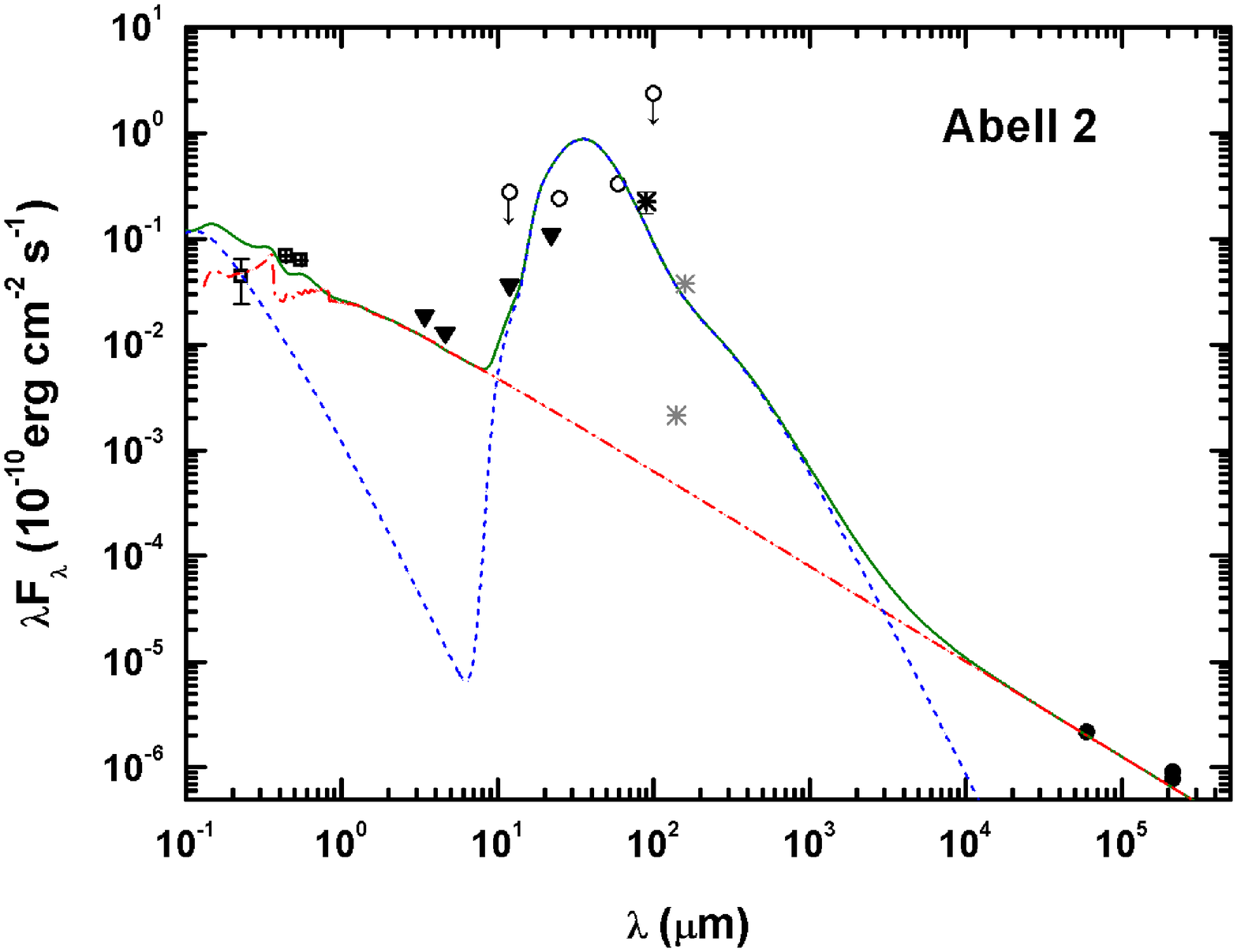}
\end{center}
\caption{The SED of Abell 2 with the spectral range from 1000~\AA~to 50 cm. The notations of data points and fitting curves are the 
same as Figure~\ref{fig6}.}
\label{fig7}
\end{figure}

To understand the properties of the ionized gas, photospheric, and dust components of these objects, the SEDs for PNs M 2-55 and Abell 2 were 
constructed (Figures~\ref{fig6} and \ref{fig7}). In the IR, the Infrared Space Observatory (ISO) Long Wavelength Spectrometer (LWS) spectrum 
was used. For photometric measurements, the near-ultraviolet (NUV) and optical B, V, i, z, and y photometry of these objects are obtained 
from \citet{Shaw85}, \citet{Martin05}, \citet{Zacharias05}, and \citet{Chambers16}. J, H, and Ks band photometric data are taken from {\it 
Two Micron All Sky Survey} (2MASS) database. In the MIR, we used the data from {\it AKARI} and {\it IRAS} Source catalogs. In addition, the 
photometric data of these PNs measured form {\it WISE} MIR images are also added. A summary of these data used is given in Table \ref{tab4}.
 
   \begin{table*}
	\caption{Photometric Measurements of M 2-55 and Abell 2}
	\centering
	\begin{tabular}{lccccc}
		\hline\hline\noalign{\smallskip}
		 & \multicolumn{2}{c}{M 2-55} && \multicolumn{2}{c}{Abell 2} \\
		 \cline{2-3}\cline{5-6} 
 Filters & Flux/Flux density & Reference && Flux/Flux density & Reference \\
		\hline\noalign{\smallskip}
		\multicolumn{6}{c}{Central star and nebula} \\
		\hline\noalign{\smallskip}
Galex NUV (mag) & ... & ... && 17.59$\pm$0.04 & \citet{Martin05} \\
B (mag) & 17.43 & \citet{Zacharias05} && 16.07$\pm$0.30 & \citet{Shaw85} \\
V (mag) & ... & ... && 15.85$\pm$0.20 & \citet{Shaw85} \\
Pan-Starrs i (mag) & 16.36$\pm$0.01 & \citet{Chambers16} && ... & ... \\
Pan-Starrs z (mag) & 16.01$\pm$0.01 & \citet{Chambers16} && ... & ... \\
Pan-Starrs y (mag) & 15.75$\pm$0.01 & \citet{Chambers16} && ... & ... \\
		\hline\noalign{\smallskip}
		\multicolumn{6}{c}{Dust$^a$} \\
		\hline\noalign{\smallskip}
2MASS J (mag) & 14.57$\pm$0.04 & \citet{Cutri03} && ... & ... \\
2MASS H (mag) & 13.75$\pm$0.05 & \citet{Cutri03} && ... & ... \\
2MASS Ks (mag) & 13.56$\pm$0.05 & \citet{Cutri03} && ... & ... \\
WISE 3.4 $\mu$m (mag) & 11.43$\pm$0.03 & this study && 12.92$\pm$0.02 & this study \\ 
WISE 4.6 $\mu$m (mag) & 10.13$\pm$0.02 & this study && 12.34$\pm$0.03 & this study \\
WISE 12 $\mu$m (mag) & 6.83$\pm$0.02 & this study && 8.35$\pm$0.04 & this study \\    
WISE 22 $\mu$m (mag) & 3.75$\pm$0.03 & this study && 5.05$\pm$0.03 & this study \\
IRAS 12 $\mu$m$^b$ (Jy) & 0.49: & \citet{Tajitsu98} && 0.11: & \citet{Tajitsu98} \\
IRAS 25 $\mu$m (Jy) & 0.80$\pm$0.07 & \citet{Tajitsu98} && 0.14$\pm$0.02 & \citet{Tajitsu98} \\
IRAS 60 $\mu$m (Jy) & 3.55$\pm$0.28 & \citet{Tajitsu98}  && 0.65$\pm$0.13 & \citet{Tajitsu98} \\
IRAS 100 $\mu$m$^b$ (Jy) & 4.99$\pm$0.40 & \citet{Tajitsu98}  && 7.77: & \citet{Tajitsu98} \\
AKARI 65 $\mu$m (Jy) & 1.85: & \citet{Yamamura10} && ... & ... \\
AKARI 90 $\mu$m (Jy) & 3.76$\pm$0.16 & \citet{Yamamura10} && 0.67$\pm$0.08 & \citet{Yamamura10} \\
AKARI 140 $\mu$m (Jy) & 3.58$\pm$0.63 & \citet{Yamamura10} && 0.01: & \citet{Yamamura10} \\
AKARI 160 $\mu$m (Jy) & 1.76: & \citet{Yamamura10} && 0.33: & \citet{Yamamura10} \\
		\hline\noalign{\smallskip}
		\multicolumn{6}{c}{Free-free emission} \\
		\hline\noalign{\smallskip}
5 GHz (mJy) & 19 & \citet{Sio01} && 2.3 & \citet{Stanghellini08} \\
4.85 GHz (mJy) & 28$\pm$4 & \citet{Gregory96} && ... & ... \\
1.4 GHz (mJy) & 26.6 & \citet{Sio01} && 7.6$\pm$0.6 & \citet{Condon98a} \\
1.4 GHz (mJy) & ... & ... && 9.4$\pm$1.1 & \citet{Condon98b} \\
		\hline\hline\noalign{\smallskip}
	\end{tabular}
	\tablecomments{0.86\textwidth}{[a] The colons represent the uncertain flux measurement. [b] Some IRAS 12 and 100 $\mu$m fluxes are 
upper limit detections.}
	\label{tab4}
\end{table*}

The preliminary impression of the SEDs (Figures~\ref{fig6} and \ref{fig7}) is that most of the fluxes from these PNs are mainly dominated in 
the IR. Assuming that the dust emission can be represented by a single blackbody (BB). The peak fluxes of these SEDs are 8.5$\times$10$^{-10}$ 
erg cm$^{-2}$ s$^{-1}$ and 1$\times$10$^{-10}$ erg cm$^{-2}$ s$^{-1}$ for M 2-55 and Abell 2, and the total fluxes emitted from the objects are 
approximately 1.2$\times$10$^{-9}$ erg cm$^{-2}$ s$^{-1}$ for M 2-55 and 1.4$\times$10$^{-10}$ erg cm$^{-2}$ s$^{-1}$ for Abell 2. Adopting 
their distances of 691 pc and 2.82 kpc for M 2-55 and Abell 2 \citep{Bailer18}, the total luminosity of these nebulae are about 18 and 35 
L$_{\sun}$ for M 2-55 and Abell 2, respectively. From Figures~\ref{fig6} and \ref{fig7}, we note that the IR SED of each PN can not be fitted 
by a single BB, thus these derived values are just the minima. 

We fitted the emerging fluxes of these objects by a two-component model including the reddened photospheric emission (emitted from the central 
star and gaseous emission) and dust continuum using the same expressions described in \citet{Hsia19}. The mid- to far-infrared component of 
the SED is not fitted by a single dust component. We have tried to fit the observed SED by a radiation transfer model using the software 
code DUSTY \citep{Ivezic99}. For the fitting, we assumed the standard MRN distribution of grain sizes \citep{Mathis77} with a dust temperature 
on inner shell boundary of 90 K and a density distribution of R$^{-2}$ for each object, where R is the distance to the central star. In 
order to arrive at the best fitting, the optical thicknesses of the shells at 0.55 $\mu$m were adopted to be $\tau$ = 3 and $\tau$ = 0.6 for 
M 2-55 and Abell 2. The dust grains are assumed to be the mixings of silicates and graphite, which can give the approximations to the dust 
continua. The separated contributions from different components of these PNs can be clearly seen in Figures~\ref{fig6} and \ref{fig7}. For the 
SED fittings, our best estimate for the central star temperatures are 80,000 K for M 2-55 and 78,000 K for Abell 2, which are in good agreement 
with previously results of 85,000 K \citep{Kaler89} and 75,000$\pm$4,000 K \citep{Shaw85} for M 2-55 and Abell 2. With the adopted distances 
of 691 pc for M 2-55 and 2.82 kpc for Abell 2 \citep{Bailer18}, the derived total luminosity of the objects are about 35 and 50 L$_{\sun}$ 
for M 2-55 and Abell 2, respectively, which are higher than our earlier estimates because parts of UV radiation emitted from the central 
stars are not counted into total observed fluxes. The higher central star temperatures and low luminosities of these PNs indicate 
that they are on the cooling paths of their evolutions in the Hertzsprung-Russell (H-R) diagram. Using the formulas presented in 
\citet{Hsia10}, we can derive the mass of dust component ($M_{d}$) and the mass of ionized gas ($M_{i}$) of PN M 2-55. Assuming the emissivity 
of dust particles of $Q_\lambda=Q_0 (\lambda/\lambda_0)^{-\alpha}$, where $Q_0=0.1$, $\alpha$ = 1, and $\lambda_0$= 1 $\mu$m. The assumed 
density of dust grain of $\rho_{s}$=1 g cm$^{-3}$ and the adopted distance of 691 pc, we obtain 4.01$\times$ 10$^{-4}$ M$_{\sun}$ and 
1.1$\times$ 10$^{-2}$ M$_{\sun}$ for the masses of the dust component ($M_{d}$) and ionized gas ($M_{i}$), respectively. The $M_{d}$/$M_{i}$ 
ratio of M 2-55 is $\sim$0.04, significantly higher than the dust-to-gas ratios of typical PNs (10$^{-2}$-10$^{-3}$, Stasi\'{n}ska \& 
Szczerba 1999). This probably suggests that a mass of gas in this evolved PN is in neutral form.  

\section{Discussion}

\subsection{M 2-55}

According to \citet{Tweedy96} and \citet{Dgani98}, the interacting PNs show three distinct features of the interactions with their 
surrounding ISM; (i) the outer regions around the PNs are asymmetric; (ii) brightness enhancements seen in the outer regions of these PNs 
accompanied by the drops in the ionization levels; (iii) the presences of fragments in the halos and/or arc-shaped filaments. 
\citet{Dgani98} suggested that the striped appearances of interacting PNs may be the result of Rayleigh-Taylor (RT) instabilities, resulting 
in fragmentations at the halos of the PNs. The optical-infrared morphology of PN M 2-55 as seen in Figures~\ref{fig1} and \ref{fig4} is 
composed of an arc-like structure with brightness enhancement on its edge from SW to SE (see Section~\ref{arc}) and an asymmetric infrared 
halo (see Section~\ref{halo}). Such features have been detected in other nebulae such as NGC 6751 \citep{Clark10}, NGC 6894 \citep{Soker97}, 
NGC 7293 \citep{Zhang12b}, HW 4, S176, and S188 \citep{Tweedy96}, and have been attributed to the interactions between the AGB winds and the 
ISM. No evidence of other two distinct characteristics of PN-ISM interactions can be seen in this object. We conclude that the presence of 
arc-shaped structure seen around PN M 2-55 is caused by the compression from the motion of this nebula through its surrounding ISM.
 
\citet{Wilkin96} found that the arc-shaped structure depends on relative velocities and densities between slow AGB wind and surrounding ISM, 
the presence of the feature can be used to determine the properties of AGB wind of the PN and its surrounding ISM. We note that the 
appearance of this new discovered arc filament is similar to that of brightening structure as the first stage of PN-ISM interaction in the 
simulations of \citet{Wareing07}, probably suggesting that the PN is not affected by the ISM interaction. If the central star of the planetary nebula (CSPN) moves slowly through the ISM (i.e. $<$50 km s$^{-1}$; Burton 1988), this interaction can keep up entire lifetime of the nebula and other PN-ISM interactions will never be observed \citep{Wareing07}. In the case of PN M 2-55, the radial velocity of this object is 22.6 
km s$^{-1}$ \citep{Acker92}, hence we infer that the arc-shaped structure around this PN may be observed for a long time.

Although M 2-55 has been known to present two pairs of bipolar lobes \citep{Sabb81}, previous images did not reveal internal details of 
this nebula. Figure~\ref{fig1} shows that the nebular structures are point symmetric about the center, and the two pairs of lobes have 
approximately the same extent, indicating that they were ejected during a short period. This PN also reveals a large central cavity, 
differing from young bipolar PNs which usually exhibit a narrow waist (see, e.g. Manchado et al. 1996; Kwok \& Su 2005). Presumably, the 
central cavity was created by the increasing thermal pressure with PN evolution. The clumpy structures in Figure~\ref{fig1} might be the 
debris of the torus that have collimated the outflows. A hypothesis of PN evolutionary transition from bipolar (lobe-dominant) to elliptical 
(cavity-dominant) has been suggested by \citet{Huarte12} and \citet{Hsia14a}. The appearance of PN M 2-55 suggests that it is in an 
evolutionary stage between the lobe-dominant phase and the cavity-dominant phase. 

\subsection{Abell 2}

According to the evolutionary tracks presented in \citet{Blocker95}, the temperature and luminosity of the central star of Abell 2 indicate that it has a mass of $\sim$0.605 M$_{\sun}$. 
This is normal among PNs \citep{Stasinska97}. \citet{Soker97a} classified Abell 2 as an elliptical PN, which might result from axisymmetric mass loss of its progenitor interacted with a substellar companion during common envelope phase. This scenario also predicts the existence of a spherical halo surrounding elliptical nebula \citep{Soker97a}, which is confirmed by our observation of Abell 2 (see Section 3.2). 

The hydrodynamical models presented by \citet{Gaw02} suggest that PNs with binary cores will evolve from bipolar to elliptical shapes. Therefore, elliptical PNs can be produced in wide binary systems and most evolved PNs appear spherical or elliptical morphologies. This arises the 
possibility that there is a companion in the nucleus of Abell 2. Although our observations could not identify whether a companion exists in this PN, new spectroscopic observation is required to reveal the nature of the nucleus in Abell 2.

The double-shell structure of Abell 2 can be explained in terms of radiation-hydrodynamics simulations \citep{Sch14}. Similar to the PN double-shell HuBi 1 \citep{Gue18}, the inner and outer shells of Abell 2 appears to have an opposite ionization structure to those of typical PNs. \citet{Gue18} suggested that the anomalous excitations are related the Wolf-Rayet central star of the PN. In this case, the nucleus of the nebula may experience a born-again event produced by a thermal pulsation at the ending of AGB stage, during the post-AGB phase \citep{Blocker01}, or on the cooling stage of white dwarf evolution \citep{Sch79}. The shocks produced by the ejecta excite the inner shell, resulting in an inverted ionization structure as seen in Abell 2.

It is instructive to investigate the evolution of SEDs form young to old PNs. The SED of a young PN IRAS 21282+5050 (hereafter referred to as 
IRAS 21282) has been presented by \citet{Hsia19}, whether a strong IR excess due to thermal emission of dust components has been revealed. In 
IRAS 21282, the dust envelope is optically thick with an optical depth of $\tau$ = 5.5 at 0.55 $\mu$m, and has a high temperature of $\sim$250 
K. It is clear that the evolved PN Abell 2 exhibits a lower optical depth and a lower temperature. Therefore, the infrared colors can reflect 
the evolutionary stage of PNs. 
  
\section{Conclusions}

Recent narrow-band imaging studies have provided an effective tool for us to understand the ionized gaseous environments around evolved PNs. 
Although more than one hundred extended, faint PNs have been studied via these advanced observational tools \citep{Tweedy96, Corradi03, 
Ali12}, the nature and properties of these nebulae are still unclear. In this paper, we present a visible and MIR study of two evolved PNs 
(M 2-55 and Abell 2). Our deep optical narrow-band images of PN M 2-55 reveal two pairs of bipolar lobes and a new arc-like structure. The 
arc-shaped structure seen around this object appears a well-defined boundary from SW to SE, furnishing a strong evidence for an interaction 
of the expanding nebula of this PN with its surrounding ISM. The [\ion{O}{III}] and H$\alpha$ images of PN Abell 2 reveal inner and outer 
elliptical shells in the main nebula, whereas the inner shell of this PN can be only seen in the [\ion{N}{II}] image. This suggests that 
the nucleus of this PN experienced a born-again event. We have studied the nebular properties of PN M 2-55 by the mid-resolution spectrum. 
The spectral analysis of this object shows that the nebula is an evolved PN with a high excitation class and a low electron density of 
250 cm$^{-3}$. 

From the MIR images of {\it WISE} all-sky survey data release, these PNs presented in our study are found to show prominent infrared features. 
Obvious MIR emissions detected in the central parts of these objects suggest that a mass of dust are located in their central regions. 
The SEDs of these PNs are constructed from extensive archival data. We successfully fitted the observed fluxes by a two-component model 
including the reddened photospheric emission and dust continuum. These PNs might have thick dusty envelopes because most of the fluxes from 
the objects emitted by dust components.

Interacting PN is thought to be the result of dynamical interaction between slow AGB wind (halo) and its surrounding ISM. The presence of 
arc-shaped/bow-shock structure suggests that PN-ISM interaction may be very common. M 2-55 and Abell 2 can serve as an astrophysical 
laboratory to study the dynamical processes of the interactions between PNs and ISM in the space.

\begin{acknowledgements}

We would like to thank the anonymous referee for his/her helpful comments that improved the manuscript. This publication has made use of data collected at Lulin Observatory, partly supported by MoST grant 108-2112-M-008-001. We acknowledge the support of the staff of the Lijiang 2.4m telescope. Funding for the telescope has been provided by Chinese Academy of Sciences and the 
People's Government of Yunnan Province. Financial support for this work is supported by the grants from The Science and Technology Development Fund, Macau SAR (file no: 061/2017/A2 and 0007/2019/A) and the Faculty Research Grants of Macau University of Science and Technology (project code: FRG-19-004-SSI). X. L. is supported by National Science Foundation of China (NSFC, grant U1731122). Y. Z thanks NSFC (NO. 11973099) for financial support.

\end{acknowledgements}

\label{lastpage}

\end{document}